\documentclass{article}

\usepackage{arxiv}

\usepackage[utf8]{inputenc}
\usepackage[T1]{fontenc}
\usepackage{hyperref}
\usepackage{url}
\usepackage{booktabs}
\usepackage{amsfonts}
\usepackage{amsmath}
\usepackage{amssymb}
\usepackage{nicefrac}
\usepackage{microtype}
\usepackage{graphicx}
\usepackage{natbib}
\usepackage{doi}

\newcommand{\coloneqq}{\mathrel{\mathop:}=}

\title{Event Horizons, Spacetime Geometry, and the Limits of Integrated Consciousness}

\author{
  Jonathon Sendall\\
  OU Philosophy Department\\
  \texttt{jonathon.sendall@ou.ac.uk}\\[0.3em]
  \href{https://jonathonsendall162367.substack.com/p/where-does-physical-reality-draw}{Click Here for the Substack Companion Article}
}

\begin{document}
\maketitle

\begin{abstract}
Extended systems are often treated as unified whenever internal causal interactions are sufficiently dense and recurrent over a characteristic timescale \citep{Tononi2004,Baars1988,Friston2010}. In relativistic spacetimes with one-way causal boundaries (such as black-hole event horizons, cosmological horizons, and Rindler wedges) and strong curvature, this background assumption can fail: closed causal loops that sustain macroscopic informational unity predicates can be obstructed without any local disruption of microphysical evolution \citep{HawkingEllis1973,Wald1984}.
This paper introduces the Reciprocal Coherence Kernel $K_T(t)$, defined as the inclusion-minimal, strongly connected operational subgraph that sustains a unified perspective according to a theory-relative functional predicate $F_T$. Kernel persistence is analysed using a relativistically explicit condition: the effective causal diameter $\Lambda(K_T(t))$ must remain less than or equal to a theory--and--model-specific coherence window $\tau_{T,m}$, measured in proper time.
When $\Lambda(K_T(t))>\tau_{T,m}$, or when a strict one-way boundary intersects the kernel, unity (as defined by the theory's own predicate) fails by coherence timeout, an operationally distinct failure mode from mechanical destruction.
A theory-relative taxonomy of geometry-sensitive versus geometry-robust unity predicates is developed. The framework is applied to horizon-straddling implementations, distributed systems, and selected fault-tolerant architectures \citep{NielsenChuang2010}.
The project is explicitly non-causal. It offers a compatibility analysis over theory-indexed model families, not a causal explanation of consciousness. The contribution lies in philosophy of science: diagnosing hidden structural commitments in integration-based theories and exposing geometric constraints on extended subjects in relativistic spacetimes \citep{Woodward2003,LadymanRoss2007}.
\end{abstract}

\section{Introduction}

\subsection{Two layers of contribution}

This paper advances two interrelated contributions, one structural and general, the other conditional and diagnostic.

First, it establishes a compatibility constraint on macroscopic unity predicates in relativistic spacetimes.
Here `theorem' is used in the structural sense: a result about theory--model compatibility under stated bridging assumptions, not an empirical discovery and not a metaphysical reduction of unity.
For any physical system whose macroscopic unity is defined in terms of reciprocal causal loops that must close within a finite coherence window $\tau_{T,m}$, global causal structure, specifically one-way causal boundaries and strong curvature, can obstruct, fragment, or time out those loops without any local disruption of microphysical evolution \citep{HawkingEllis1973,Wald1984,Earman1995}.

Second, the constraint is applied conditionally to a family of scientific theories that characterise the unity of consciousness (or other macroscopic informational unities) in terms of reciprocal closure.
Integrated Information Theory (IIT), Global Workspace Theory (GWT), Predictive Processing (PP), and some engineered recurrent-control architectures share a structural commitment: unity requires the existence of a reciprocally interacting structure that is functionally adequate on the theory's own criteria \citep{Tononi2004,TononiEtAl2016,DehaeneChangeux2011,MashourEtAl2020,Friston2010,Clark2013}.
If a target theory $T$ identifies experiential or operational unity with the presence of such a structure, then the geometric constraints derived here restrict $T$'s admissible unity claims in spacetimes containing event horizons, Rindler wedges, or cosmological horizons \citep{Malament2012}.

Throughout, `unity' is shorthand for a theory's own unity predicate.
The constraints apply to unity predicates and their model families, not to the metaphysical nature of unity as such \citep{Bayne2010,Chalmers2010}.

\subsection{Bridge principle}

\medskip
\noindent
\textbf{Bridge Principle.}
If a theory $T$ treats the unity of a conscious field (or other macroscopic informational subject) as requiring an operationally diagnosable form of reciprocal causal closure over a characteristic coherence timescale $\tau_{T,m}$, then the geometric results derived here constrain $T$'s unity claims for any model class $m$ within its intended scope of application, including analogue and simulation regimes where its diagnostic quantities are well-defined \citep{Woodward2003,FriggHartmann2020}.
\medskip

The argument does not endorse any particular theory $T$, and it does not propose a new positive theory of consciousness.
It analyses how spacetime geometry functions as an external filter on theories that already tie unity to finite-window reciprocal integration.

\subsection{Ontology vs epistemology of unity}

The framework adopts a two-level stance.

At the epistemic or operational level, kernels and persistence conditions are modelling tools.
They formalise which parts of a system can bidirectionally influence one another within $\tau_{T,m}$, respecting relativistic signal limits \citep{Woodward2003,Hitchcock2001}.

At the ontic level, the framework remains neutral.
It does not claim that being a kernel is metaphysically constitutive of unity.
It claims that any account that uses reciprocity-within-$\tau_{T,m}$ as its diagnostic mark inherits geometric constraints, whether reciprocity is treated as constitutive or merely evidential \citep{Bayne2010,Parfit1984}.

In this restricted operational sense, kernel cessation may be glossed as a `pseudo-death' of the $T$-subject role, meaning failure of $T$'s own unity diagnostics (coherence timeout), not a claim about metaphysical persistence or personal identity.
This label is optional and purely expository, and it does not settle questions of personal identity or survival.

\subsection{One-way causal boundaries as intervention constraints}

One-way causal boundaries are characterised interventionally.
Region $\mathcal{B}$ lies behind a one-way boundary relative to region $\mathcal{A}$ (within $\tau_{T,m}$) if no intervention performed in $\mathcal{B}$ can have an effect in $\mathcal{A}$ that is available for feedback to $\mathcal{B}$ within $\tau_{T,m}$, while interventions in $\mathcal{A}$ can still produce effects in $\mathcal{B}$ within $\tau_{T,m}$.

In black-hole spacetimes, once a worldline crosses the event horizon, there are no future-directed causal curves from interior events to exterior events \citep{HawkingEllis1973,Wald1984,Malament2012}.
That condition precludes return-directed intervention and therefore precludes any kernel whose unity predicate requires reciprocal closure across the boundary.

It is useful to distinguish strict one-way boundaries from effective one-way boundaries.
Strict one-way boundaries prohibit interior-to-exterior causal curves (event horizons and cosmological horizons) \citep{HawkingEllis1973,Wald1984}.
Effective one-way boundaries permit return curves but fail to support feedback closure within $\tau_{T,m}$ under bandwidth, noise, delay, or processing constraints \citep{SterbenzEtAl2010,HoneyEtAl2007}.

\subsection{Methodological note}

\textit{Open question (orientation).}
For predicates whose $F_T$ depends on particular $F_T$-essential loops, the order in which a one-way boundary intersects those loops can affect which kernels remain persistence-satisfying.
An \textit{orientation profile} is the temporal ordering of boundary intersections with $F_T$-essential loops in $G(S,t,\tau_{T,m})$; any neuroanatomical cases (hemisphere first, callosal first, spinal first) function here only as schematic illustrations of different profiles, not as empirical claims about brain structure.

\section{The Reciprocal Coherence Kernel}

To move beyond global graph connectivity, we distinguish broad causal structure from the minimal functional architecture required to sustain a unified perspective \citep{Sporns2013,BullmoreSporns2009}.

In what follows, terms such as `capacity for a perspective' and `operational subject' are functional placeholders.
They name whatever role a given theory $T$ assigns to the structures it treats as realisers of unity in model class $m$ \citep{Dennett1991,Bayne2010}.

\subsection{Functional selection schema}

Let $S$ be a physical system composed of individuated parts (nodes) $\{n_i\}$ evolving along worldlines in a spacetime manifold $(M,g_{ab})$.
Let $G(S,t,\tau_{T,m})$ be the effective causal graph of the system over an interval $[t,t+\tau_{T,m}]$.
Edges represent physically realised causal influences.
A directed edge $A\to B$ indicates that $A$ can causally influence $B$ via future-directed causal curves and internal signalling channels within the coherence window \citep{Pearl2000,Woodward2003}.

Introduce a functional selection predicate $F_T(\cdot)$ parameterised by theory $T$.
A subgraph $G'\subseteq G(S,t,\tau_{T,m})$ satisfies $F_T(G')$ if and only if $G'$ contains sufficient structure to instantiate whatever $T$ counts as a unified self-frame or operational subject in model class $m$ \citep{Tononi2004,DehaeneChangeux2011,Friston2010}.

The point is not to claim that $F_T$ is easy to compute.
The point is to isolate where the geometric constraint enters once a theory has committed itself to a reciprocity-based diagnostic.

\subsection{Kernel definition and degeneracy}

The Reciprocal Coherence Kernel is defined as an inclusion-minimal strongly connected component (SCC) that satisfies $F_T$:
\[
K_T(t)\;=\;\min_{\subseteq}\Big\{\,G' \subseteq G(S,t,\tau_{T,m})\;\Big|\;G' \text{ is an SCC and } F_T(G')\,\Big\}.
\]
Minimality means no proper subset of $K_T(t)$ suffices to satisfy the unity predicate under $T$.
A loop is essential relative to $F_T$ if ablation of that loop causes $F_T$ to fail when evaluated using the theory's own diagnostic quantities.

Let $\mathcal{K}_T(t)$ be the set of all inclusion-minimal kernels at time $t$.
Define the kernel degeneracy index as
\[
D_T(t;m,S)\;\coloneqq\;\big|\mathcal{K}_T(t)\big|.
\]
High $D_T$ indicates predicate underdetermination for that theory--model pair.
This is a structural tool, not a metaphysical conclusion \citep{Quine1975,LadymanRoss2007}.

\subsection{Relativistic operational bound: kernel persistence}

Instead of a spatial diameter, use an operational round-trip latency.

Fix a reference worldline $\gamma$ co-moving with the kernel, with proper time as the clock convention.
Let $\tau_{T,m}$ be the coherence window, measured in that proper time.
For nodes $A,B\in K_T(t)$, let $T_{A\to B\to A}$ be the infimum proper time on $A$'s clock between sending a signal to $B$ and receiving a return signal, if any return-directed causal curve exists.\footnote{For compact systems in weak-field regimes (e.g.\ biological brains), proper-time differences between $\gamma$ and individual nodes are negligible. For spatially extended systems near strong-curvature regions, relative time dilation may cause node clocks to diverge from $\gamma$; in such cases, $\tau_{T,m}$ should be understood as a constraint on $A$'s local proper time, or else rescaled appropriately for the reference frame in use.}
If no such return-directed causal curve exists, define $T_{A\to B\to A}\coloneqq\infty$.

Define the effective causal diameter
\[
\Lambda\!\left(K_T(t)\right)\;=\;\sup_{A,B\in K_T(t)}\Big\{\,T_{A\to B\to A}\,\Big\}.
\]
The Kernel Persistence Condition is
\[
\Lambda\!\left(K_T(t)\right)\;\le\;\tau_{T,m}.
\]

For strict one-way boundaries, obstruction is qualitative.
If a boundary removes all return-directed causal curves for at least one required loop, then $\Lambda=\infty$ and persistence fails for any finite $\tau_{T,m}$.

For effective one-way boundaries, return curves exist but become unusable under delay, noise, or bandwidth constraints.
In that case, $\Lambda$ can increase continuously until it exceeds $\tau_{T,m}$.

Define the \textit{kernel-cessation time} $t^{*}$ as
\[
t^{*}\;\coloneqq\;\inf\Big\{\,t\;:\;\nexists\,K_T(t)\ \text{such that}\ K_T(t)\ \text{satisfies}\ F_T\ \text{and}\ \Lambda(K_T(t))\le\tau_{T,m}\,\Big\}.
\]
If $D_T(m,S) > 1$, cessation occurs when the last persistence-satisfying kernel fails the bound, rather than when any particular kernel fails.
In strict one-way boundary cases this coincides with the first time the kernel intersects the boundary, since $\Lambda(K_T(t))=\infty$ then violates any finite $\tau_{T,m}$.
This is an operational failure of unity-attribution under $T$'s diagnostics, not a claim about substrate destruction or metaphysical persistence.

\section{Event horizons as strict one-way boundaries}

This section shows how to stress-test kernel persistence under a strict one-way boundary.
The SCC fact remains true, but it is now used as a lemma inside a more diagnostic claim about which unity predicates fail, and why.

\subsection{Cross-horizon fragmentation with kernel interpretation}

\begin{figure}[t]
\centering
\includegraphics[width=0.85\linewidth]{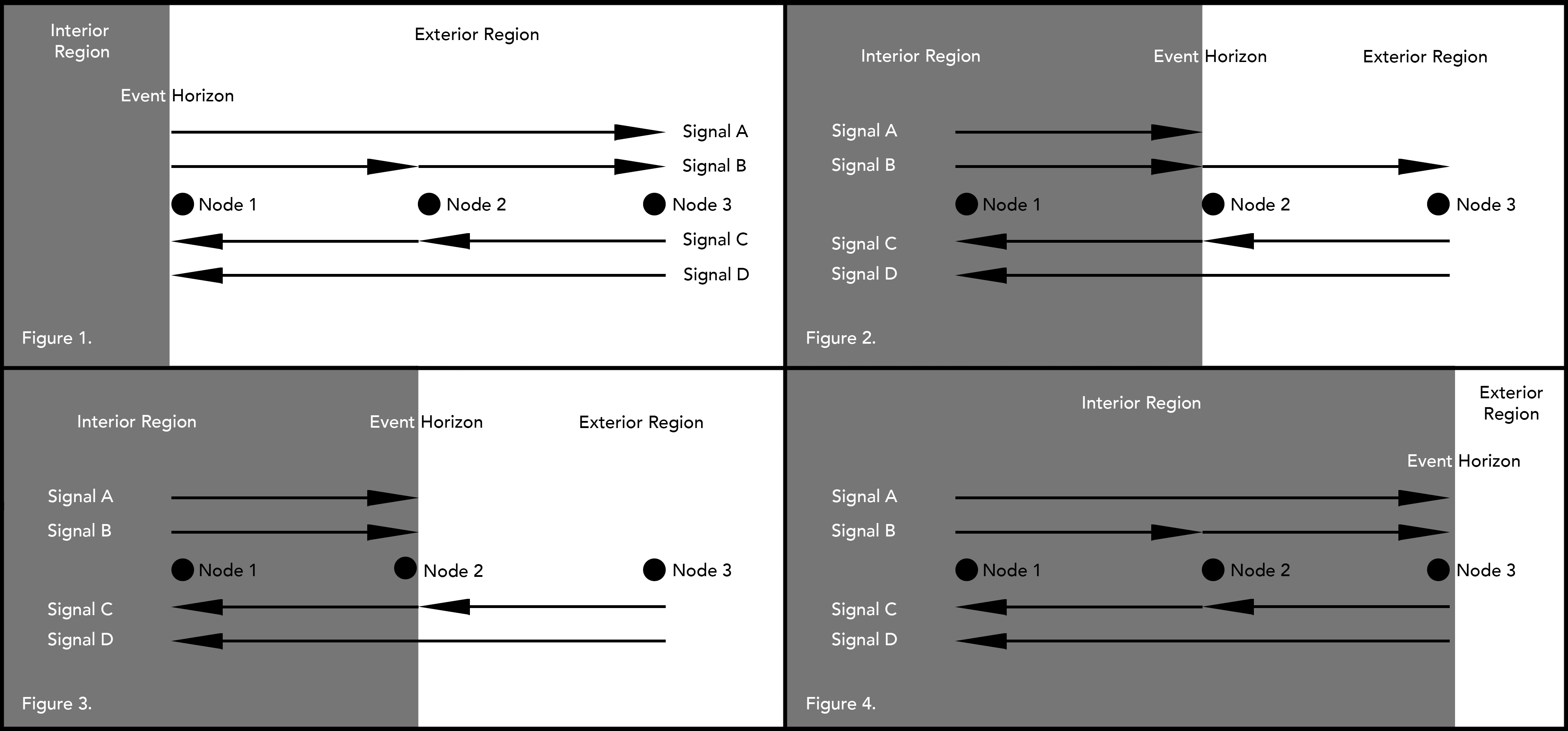}
\caption{Causal fragmentation of a recurrent three-node system (N1--N2--N3) in radial free-fall crossing a Schwarzschild event horizon. N1 is innermost (left), N3 outermost (right). Outward signals: A propagates directly from N1 to N3; B(1) and B(2) propagate from N1 to N3 via N2. Inward signals: D propagates directly from N3 to N1; C(1) and C(2) propagate from N3 to N1 via N2. As nodes cross the horizon, outward signals are blocked at the horizon while inward signals continue, breaking strong connectivity during the straddling phase.}
\label{fig:fragmentation}
\end{figure}

Figure~\ref{fig:fragmentation} illustrates the causal-fragmentation pattern for a recurrent three-node system crossing a Schwarzschild event horizon.

In panel 1 (top-left), all three nodes lie in the exterior.
Outward signals A, B(1), and B(2) propagate from Node 1 toward Node 3.
Inward signals C(1), C(2), and D return from Node 3 toward Node 1.
All loops close within $\tau_{T,m}$, so there exists an SCC spanning N1--N2--N3.
If $F_T$ selects this recurrent triad as functionally adequate, then a kernel may coincide with, or be contained in, that SCC.

In panel 2 (top-right), Node 1 has crossed the horizon while Node 2 remains outside near the boundary.
Signals A and B(1), originating from Node 1, are blocked at the horizon and cannot reach Node 2 or Node 3.
Signal B(2) may still propagate from Node 2 to Node 3.
Inward signals from Node 3 still cross the horizon and reach Nodes 1 and 2.
Strong connectivity across all three nodes has failed.
Kernel interpretation: any unity predicate whose essential loops require interior-to-exterior return paths loses the ability to satisfy $F_T$ on a kernel that includes both sides.

In panel 3 (bottom-left), Nodes 1 and 2 are inside while Node 3 remains outside.
Outward signals A and B(1) are blocked, and B(2) is blocked at source.
Inward signals C(1), C(2), and D still cross the horizon.
Node 3 receives no return signals, so no bidirectional exchange is possible.
Kernel interpretation: the exterior region may retain a trivial SCC, but it cannot support any kernel whose $F_T$ requires a non-trivial recurrent core, so an exterior unity predicate can fail by functional inadequacy without any local damage.

In panel 4 (bottom-right), all three nodes occupy the interior.
Strong connectivity is restored among them, since all signals can again propagate in both directions within the interior.
Kernel interpretation: interior unity can reconstitute because the interior domain still permits return-directed causal curves among interior nodes.
This interior kernel is cut off from exterior observers.

\subsection{Proposition: no kernel can straddle a strict one-way boundary}

Let $\mathcal{H}$ be a strict one-way boundary relative to an exterior region.
Suppose $F_T$ requires, as an essential condition, at least one return-directed influence from an interior node to an exterior node within $\tau_{T,m}$ for the kernel under consideration.
Then no kernel satisfying $F_T$ can include nodes on both sides of $\mathcal{H}$ during any interval in which the system straddles $\mathcal{H}$.

The proof is immediate from strict one-way structure.
Interior-to-exterior return-directed causal curves do not exist \citep{HawkingEllis1973,Wald1984}.
Therefore at least one required loop has $T_{A\to B\to A}=\infty$, so $\Lambda(K_T(t))=\infty$, which violates $\Lambda(K_T(t))\le\tau_{T,m}$ for any finite $\tau_{T,m}$.

The claim is restricted to those unity predicates that actually require cross-boundary return paths as part of their minimal functional core.

\subsection{Timeout before crossing}

In strict boundary cases, straddling produces an abrupt structural disconnection.
In addition, in pre-crossing regimes, redshift and processing delays can make return loops unusable within $\tau_{T,m}$ even before geometric crossing for architectures whose essential loops are radially extended.

This isolates the question that matters operationally.
A theory can be geometry-sensitive even if its kernels do not literally straddle a horizon, because its essential loops can exceed the coherence budget in the approach region.

\section{Applications}

\subsection{Horizon-straddling subjects}

The structural takeaway is diagnostic rather than melodramatic.
A horizon does not force a single metaphysical story about persons.
It forces a constraint on reciprocity-based unity predicates.
If a theory requires return signalling within a finite window for unity, the straddling phase removes the possibility of satisfying that requirement across the boundary.

Post-fragmentation `reintegration' can mean distinct structural outcomes:
\begin{enumerate}
\item \textbf{Re-expansion}: a persistence-satisfying kernel remains non-empty throughout the straddling interval and later grows as geometry permits.
\item \textbf{Fusion (degeneracy reduction)}: $D_T(m,S) > 1$ during an interval but later returns to $1$ via merger within a region where reciprocal closure is again available. This is a non-strict-boundary possibility; strict horizons preclude cross-boundary fusion.
\item \textbf{Reconstitution after a gap}: no persistence-satisfying kernel exists for some interval and one later forms.
\end{enumerate}
These labels track kernel structure under $T$ and do not adjudicate identity claims.

Later $D_T(m,S)=1$ does not entail $D_T(m,S)=1$ throughout; it blocks only the backward inference from later unity to earlier non-multiplicity.
Reintegration does not deny multiplicity when $D_T(m,S)>1$ during the straddling regime.
It denies only the inference from later unity to earlier uniqueness.

\subsection{Distributed systems and effective one-way boundaries}

The same formalism applies when the boundary is not geometric but operational.
A network can contain channels that are so delayed, lossy, or bandwidth-limited that return loops cannot close within $\tau_{T,m}$.
In that case, unity can fail by timeout even though return curves exist in principle \citep{SterbenzEtAl2010}.
This produces an effective analogue of the strict-horizon case.

\subsection{Quantum error-correcting and fault-tolerant architectures}

Some fault-tolerant schemes preserve logical information under local noise.
That does not imply they preserve the reciprocal causal structure required by a given unity predicate.
If the unity predicate relies on classical feedback cycles across a partition, a strict one-way boundary still blocks those cycles, even if quantum correlations persist \citep{NielsenChuang2010}.
The framework is therefore a filter on which unity predicates can be carried by which architectures under which causal embeddings.

\section{Unity taxonomy as idealisation ladder}

The kernel framework yields a taxonomy of unity predicates distinguished by sensitivity to global causal structure.
This taxonomy classifies predicates and their model families, not unity in itself \citep{Bayne2010}.

\begin{itemize}
\item \textbf{Type I: Strongly geometry-sensitive (distributed).} Every admissible kernel includes essential loops whose return paths fail under strict one-way boundaries. Straddling forces kernel failure or fragmentation.
\item \textbf{Type II: Moderately geometry-sensitive (pressure-sensitive).} Kernels can be compact enough to survive locally, but typical implementations rely on extended loops. Unity can persist briefly via kernel retreat, then fail by timeout under coherence pressure.
\item \textbf{Type III: Geometry-robust (effectively local relative to $\tau_{T,m}$).} Kernels remain effectively local, with $\Lambda(K_T)\ll\tau_{T,m}$ in all relevant configurations. Unity fails only by direct local damage or singularity encounters.
\end{itemize}

This ladder preserves the geometric insight while restricting the strongest bifurcation-style rhetoric to Type I predicates.

\section{Open questions and wider implications}

The kernel framework and its application to one-way boundaries raise several questions that extend beyond the immediate technical results.
This section flags these as open problems rather than resolved theses; a full treatment would require engagement with literatures in philosophy of mind, personal identity, and foundations of physics that lie outside the present scope.

\subsection{Geometric cloning of unity predicates}

During the straddling phase, the degeneracy index $D_T(m,S)$ can exceed unity: two (or more) disjoint kernels, each satisfying $F_T$, may coexist simultaneously.
If the target theory $T$ treats $F_T$-satisfaction as sufficient for unified subjecthood, then the same substrate that previously supported one $T$-subject now supports two.

This is not duplication of matter but duplication of the causal-closure condition that $T$ requires for unity.
The question this raises is whether `cloning' of a unity predicate by geometric means has the same implications as cloning by substrate duplication.
Standard fission scenarios in the personal-identity literature involve copying or splitting the physical realiser \citep{Parfit1984,Shoemaker1984}.
The horizon case holds the realiser fixed and varies only the geometry.
Whether this difference matters for identity claims is not settled by the present framework, but the framework makes the question precise: under what conditions does $D_T > 1$ during an interval $[t_1, t_2]$ entail subject-multiplication in whatever sense the theory $T$ endorses?

\subsection{Substrate-only supervenience}

A standard physicalist commitment holds that fixing all physical facts about a system fixes all mental facts about that system \citep{Kim1998,Chalmers2010}.
The horizon case suggests a refinement.
Two systems with identical local physical states, identical functional organisation, and identical internal causal capacities can differ in $D_T$ solely due to differences in their spacetime embedding.
If $D_T$ tracks the number of $T$-subjects, then subject-count does not supervene on substrate alone; it supervenes on substrate-plus-geometry.

This does not refute physicalism, but it constrains its formulation.
The supervenience base for unity predicates, on any reciprocity-based theory, must include the global causal structure of the region the substrate occupies, not merely the intrinsic properties of the substrate itself.
The horizon makes this visible; ordinary spacetime conceals it by satisfying the geometric preconditions so abundantly that they appear trivial.

\subsection{A maximum spatial extent for unified fields}

Even in flat spacetime, the Kernel Persistence Condition $\Lambda(K_T(t)) \le \tau_{T,m}$ imposes a ceiling on the spatial extent of any unified field.
For a system in Minkowski spacetime with characteristic coherence window $\tau_{T,m}$, no kernel can have effective causal diameter exceeding $c\tau_{T,m}$.
For biological timescales ($\tau \sim 100$--$500\,\mathrm{ms}$), this ceiling is of order $10^7$--$10^8\,\mathrm{m}$, far larger than any terrestrial system.

For distributed artificial systems, the constraint becomes non-trivial.
A network spanning light-seconds inherits the same geometry-dependent individuation that the horizon case dramatises, even without curvature or strict boundaries.
The question this raises is whether the individuation of large-scale information-processing systems, including hypothetical interplanetary or interstellar architectures, is subject to hard geometric limits that current engineering frameworks do not explicitly address.

\subsection{Death by dimensional exclusion}

The exterior kernel during straddling can shrink below the threshold of $F_T$-adequacy without any local damage to the substrate.
Neurons continue firing; processing units remain powered; no lesion or hardware failure occurs.
What ends is the geometric possibility of reciprocal closure across the kernel's former extent.

This constitutes a failure mode distinct from both mechanical destruction and local functional degradation: call it \textit{cessation by dimensional exclusion}.
The parts persist; the unity predicate fails.
Whether this amounts to `death' of the $T$-subject depends on whether $T$ individuates subjects by substrate or by kernel structure.
The framework does not answer this question, but it isolates the structural distinction that any answer must address.

\subsection{Reframing the hard problem}

The traditional hard problem asks how physical processes give rise to qualitative experience \citep{Chalmers1996}.
The horizon case suggests a prior question: what geometric conditions must a region satisfy for the causal closures required by experience to be possible at all?

On reciprocity-based theories, unity requires closed loops within $\tau_{T,m}$.
The existence of such loops is not guaranteed by substrate properties; it depends on the metric structure of the ambient spacetime.
Ordinary spacetime supplies this structure so generously that the dependence is invisible.
The horizon reveals it by withdrawing the structure locally.

This does not dissolve the hard problem, but it suggests that the problem may have been incompletely framed.
The explanatory gap may not lie solely between physical processes and phenomenal properties; it may also involve the geometric preconditions under which those processes can constitute a unified subject.
A full treatment of this suggestion would require resources from philosophy of mind that the present paper does not deploy, but the kernel framework provides the structural vocabulary in which the question can be posed.

\subsection{Implications for the equivalence principle}

The equivalence principle guarantees that local physics along a freely falling worldline is indistinguishable from local physics in flat spacetime \citep{Wald1984}.
The horizon argument does not violate this principle: no local anomaly occurs at the horizon.
What fails is global reciprocal closure, a property that depends on the light-cone structure of extended regions rather than on local dynamics.

This raises a question about the scope of `local supervenience' claims for consciousness.
If experiential unity supervenes on local physical states, then bifurcation at the horizon would be inexplicable: local states are continuous across the crossing.
The framework suggests that unity predicates with essential loops extending beyond a single worldline cannot supervene on local states alone; they inherit sensitivity to global causal structure.
This is a constraint on the form that supervenience claims can take, not a violation of any physical principle.

\section{Conclusion}

This paper developed a geometry-dependent framework for analysing macroscopic informational unity predicates in relativistic spacetimes.
The central construct, the Reciprocal Coherence Kernel $K_T(t)$, is defined via a theory-relative functional predicate $F_T$ and constrained by the Kernel Persistence Condition $\Lambda(K_T(t)) \le \tau_{T,m}$.
The framework distinguishes global fragmentation from kernel failure, and distinguishes coherence timeout from mechanical destruction.

The result is a philosophy-of-science contribution: it diagnoses which hidden structural commitments integration-style unity predicates carry, and it shows how strict and effective one-way boundaries constrain unity claims without adding a new causal story about consciousness \citep{Woodward2003,FriggHartmann2020}.

The open questions flagged in Section 6 indicate that the geometric constraints derived here may have implications for debates in philosophy of mind, personal identity, and foundations of physics.
These implications are not argued to conclusion in the present paper; they are offered as problems that the kernel framework makes precise and that subsequent work might address.

These results constrain theory-relative unity attribution under $T$'s diagnostics; questions of personal identity or survival remain underdetermined by kernel facts alone.

\end{document}